# Investigation of geometry-dependent field properties of 3D printed metallic photonic crystals

**Dejun Liu,[1,2, *] Siqi Zhao,[1] Borwen You,[1] Toshiaki Hattori,[1, *]**

[1]*Department of Applied Physics, University of Tsukuba, Tsukuba, 305-8573, Japan*
[2]*Department of Physics, Shanghai Normal University, Shanghai 200234, China*
*\*Corresponding author: dejunliu1990@gmail.com, hattori@bk.tsukuba.ac.jp*

**One terahertz (THz) waveguide based on 3D printed metallic photonic crystals is experimentally and numerically demonstrated in 0.1–0.6 THz, which consists of periodic metal rod arrays (MRAs). Results demonstrated that such waveguide supports two waveguide modes such as fundamental and high-order TM-modes. The high-order TM-mode shows high field confinement, and it is sensitive to the geometry changes. By tuning the metal rod interspace, the spectral positions, bandwidths, and transmittances of the high-frequency band can be optimized. The investigation shows that a mode conversion between high-order modes occurs when the MRAs symmetry is broken via change the air interspace.** © 2020 Optical Society of America

Recently, optical waveguides are important for terahertz (THz) applications in communications, sensing, and imaging [1-2]. In contrast to conventional fibers, free-standing photonic crystals (PCs) based waveguides are attractive alternatives that offer further advantages such as compact size and broad stopband [3-5]. Different from dielectric, metal exhibit huge conductivity in THz range [6-8]. Thus, the metal-air device with low-loss and low-dispersion was intensively studied for waveguide purposes [9-11]. Owing to the exists surface plasmon polaritons (SPPs), metal photonic crystals enable to confine the THz wave in the sub-wavelength scale [12-20]. A photonic crystal based on metal rod array (MRA) has been experimentally demonstrated in 0.1–1 THz, which shows a low waveguide propagation loss of 0.03 cm$^{-1}$ and high field confinement [19]. By optimizing structural geometry, the loss and field confinement of MRAs can be improved. Theoretical results demonstrated that arranging the interspace between metal rods is a critical stratagem to optimize the transportation efficiency of THz waves through an MRA structure [20]. However, the experimental work of geometry-dependent field properties of MRAs has yet been explored because it is still a challenge to fabricate a miniature photonic crystal device in THz [21]. Conventional electron beam lithography and direct laser writing are limited in their capabilities to produce structures in miniature []. The procedures of lithography are complicated and time-consuming, which also need an additional mask and etch chemical materials [19]. Fortunately, a relatively simple fabrication of 3D printing with low-cost and timeless has been proposed. 3D printer based on stereolithography (SLA) with high accuracy are widely used to fabricate THz waveguides [22-24].

Here, we studied a 3D printed metallic photonic crystal based on metal rod arrays (MRAs) in THz frequencies. As we have discussed in our previous studies [20], by changing the metal rod interspace, the spectral positions, transmittance, and bandwidth can be optimized. To prove the simulation results, several MRAs with various metal rod interspaces are experimentally demonstrated.

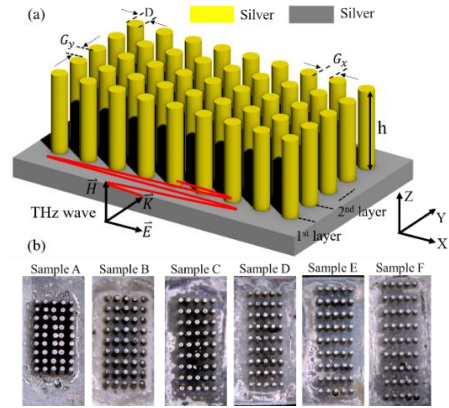

| Sample No. | $G_x$ (mm) | $G_y$ (mm) | $h$ (mm) | $D$ (mm) | $\Lambda_y$ (mm) |
|---|---|---|---|---|---|
| A | **0.252** | 0.291 | 1.775 | 0.330 | 0.621 |
| B | **0.318** | 0.162 | 2.046 | 0.394 | 0.556 |
| C | **0.587** | 0.367 | 2.034 | 0.228 | 0.595 |
| D | **0.644** | 0.295 | 2.075 | 0.279 | 0.574 |
| E | **0.722** | 0.296 | 2.050 | 0.309 | 0.595 |
| F | **0.815** | 0.292 | 2.013 | 0.283 | 0.576 |

**Fig. 1.** (a) The configuration of a 3D printed metal rod array. (b) (b), the microscopic photos for MRAs with various $G_x$. Table: The measurement structure parameter for various MRAs

The MRA model configuration is schematically depicted in Fig. 1 (a). The MRA consists of a periodically arranged uniform metal rod of 9 lines and 5 layers. The metal rod has an identified height *h* and diameter *D*. The air interspace between rods in the X- and Y-axis are defined as $G_x$ and $G_y$, respectively. Thus, the period of the MRA ($\Lambda$) is determined from the rod diameter *D* and the air interspace *G*. The top-view photos of fabricated MRAs shown in Fig. 1 (c). The fabrication process of the device is described as follows. Firstly, a resin (405 nm UV-resin) structure is fabricated using a 3D printer (ANYCUBIC Photon, UV-LED-405 nm) with a transverse resolution of 47 μm and a longitudinal resolution of 1.25 μm (i.e., along the structure height). After the resin structure is developed in the 99% ethanol and solidified by a UV LED, the polymer rod array (PRA) is obtained. The polymer rod array is then coated with a silver metal thickness about 200 nm, using a sputter coating system (CFS-4EP-LL), to obtain the metal rod array (MRA). The silver thickness of 200 nm is larger than the skin depth of 1 THz waves, which is around 100 nm. The table in Fig. 1 shows the measured structural parameters for several MRAs with various $G_x$. Only $G_x$ values larger than 0.25 mm are fabricated due to the limited accuracy of the 3D printer.

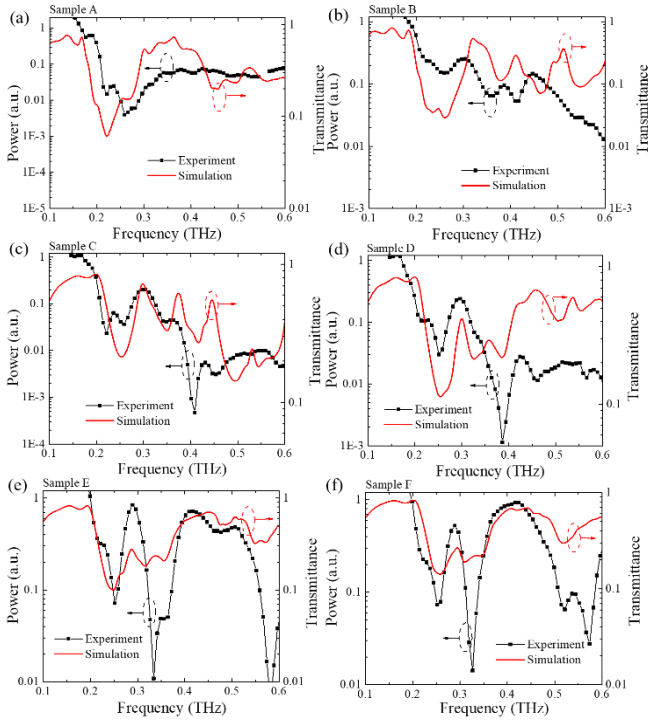

**Fig. 2.** The experimental and simulated transmission spectra of MRAs with various $G_x$.

The waveforms of the transmitted THz waves along the metal rod array at edge-coupling can be reliably measured by a terahertz time-domain spectroscopy (THz-TDS) based on Ti: Sapphire laser (780 nm). The measurement spectra of samples A, B, C, D, E, and F are shown in Fig. 2 (a-f), respectively. Experimental and calculated results proven that a noticeable Bragg reflection band occurs along the MRAs, which is found in the photonic crystal [5, 19-21]. Indicated that 5 layers of MRA behave as a THz photonic crystal. The experimental results agree well with that of simulated results using 3D-FDTD in which the little discrepancy between experimental and simulated transmission spectra results from the un-uniform rods and rough metal surface. The bandwidth and center frequency of bandgap is determined from the diameter and period of rods. For example, sample A realizes a noticeable bandgap width of 60 GHz with a center frequency of 0.22 THz because of the narrow air interspace ($G_x$=0.252 mm). With the increases $G_x$, from 0.252 mm to 0.815 mm, the high-frequency band shows different variations. Compared with sample A, the high-frequency band of sample F shows a transmission dip at 0.33 THz. It means that larger $G_x$ MRA results in two transmission bandgaps. These results suggest that the number of transmission bands can be manipulated by changing $G_x$ [20].

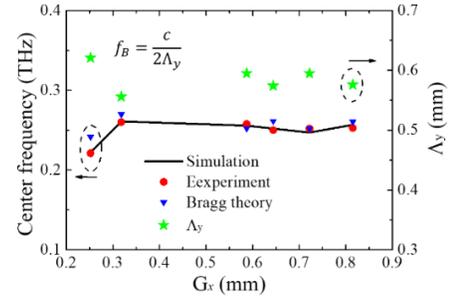

**Fig. 3.** The center frequency of 1st bandgap with the varying of $G_x$ and $\Lambda_y$.

To further know the relation between the center frequency of first (1st) bandgap and the metal rod interspace $G_x$, we summarized these results and shown in Fig. 3. Experiment results show good agreement with Bragg theory and simulation results. The small discrepancy of bandgap center frequency between the 3D-FDTD and Bragg theory results from the different structural boundary and input THz beam sizes. For 3D-FDTD, the input THz beam approximating to the rod length about 2 mm. However, in Bragg theory, the boundary is 2D periodic boundary with an infinite rod length, and the input beam considers as a plane wave cover the MRA cross-section area. The experimental result agrees well with that of simulation because the input beam in 3D-FDTD is a Gaussian beam approximating THz beam in the experiment (the red-circle dot and black line). As discussed in previous studies [20], the bandgap center frequency is tightly correlated with the air interspace of MRA in the propagation direction ($\Lambda_y$) rather than that of $G_x$. Results proved that with the increases $G_x$, the center frequency of 1st bandgap has no obvious shifts. The center bandgap frequency of sample F ($G_x$ =0.815 mm) is 0.25 THz, which is calculated by equation $f_B = c/2(D + G_y)$.

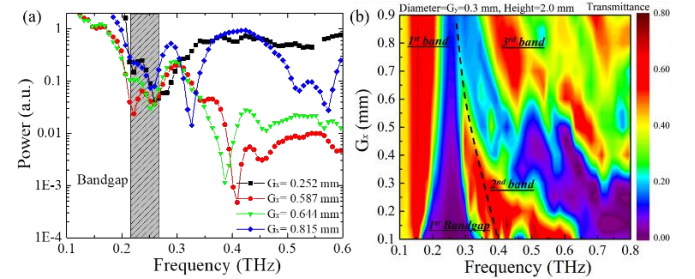

**Fig. 4.** (a) Measured spectra of MRAs. (b) The counter map of calculated transmission spectra of MRAs with different $G_x$ values (0.1-0.9 mm), where the rod diameter, height, and $G_y$ are, respectively, 0.3 mm, 2.0 mm and 0.3 mm.

The changes of $G_x$ cause the shifts of the MRA TM modes, which have been proved in our previous simulation works [20]. Here, four MRAs with $G_x$ of 0.252, 0.587, 0.644, and 0.815 mm are selected to study. As shown in Fig. 4 (a), an obvious bandgap can be found in 0.2-0.3 THz. For the increased $G_x$, the bandwidth of 1st bandgap becomes narrower. It indicates that the large $G_x$ values weaken the destructive interference effect among the metal rods. Furthermore, the first pass-band shifts to the higher frequencies and its transmittance are increased. The transmittance trend of the 1st pass-band peak is proportional to $G_x$, which is consistent to that of one PPWG that depends on its hollow core space [9-10]. Conversely, the high-order modes in the second pass-band of 0.3-0.4 THz exhibit a redshift with the increased $G_x$. Noted that the cutoff frequencies of the high-order mode are inversely proportional to the $G_x$ values, which can be expressed as $v_c = mC/2n_{eff}G_x$, where $v_c$, $m$, $C$, and $n_{eff}$ are the cutoff frequency, number of TM waveguide mode, light speed in vacuum, and effective waveguide refractive index, respectively [20]. For 0.252 mm $G_x$-MRA, the spectral peak locates at 0.366 THz. When $G_x$ up to 0.815 mm, its frequency is reduced from 0.366 to 0.293 THz. Besides, the transmittance of that spectral peak decreases from 0.72 to 0.23. Figure 4 (b) shows the counter map of the calculated transmission spectra of MRAs with different $G_x$. In the simulation, the rod diameter, height, and $G_y$ are fixed as 0.3, 2.0 and 0.3 mm, respectively. The 2nd and 3rd band are shifted to the lower frequencies. In other words, the cutoff frequencies of the 2nd and 3rd modes are reduced with the increased $G_x$. The waveguide-mode bandwidths based on the full width at half maximum (FWHM) and the corresponding peak transmittance that depends on $G_x$ are further observed. For the increased $G_x$, the 2nd band is shifted to the low-frequencies; its FWHM and transmittance are clearly decreased, which agrees well with experimental results in Fig. 4 (a). These results show that the high distinction of lateral confinement and waveguide transmittance at the 2nd TM mode spectrum is consequently determined by the asymmetric interspace of $G_x$. However, in contrast to the 2nd band, the transmittance of the 3rd band is improved for the $G_x$ increment.

the electric field distribution respectively at 0.170, 0.293, and 0.424 THz for 0.815 mm- $G_x$ MRAs.

Figure 5 demonstrates the $G_x$-dependent modal field distributions of three bands at the corresponding spectral peak. For 0.252 mm-$G_x$ MRAs, the 0.168 THz wave is tightly confined at the top MRA regions. It suggests that low-frequency THz waves can be confined and to be guided in a narrow interspace of MRA. This result coincides with that of the single metal slit [94]. The 0.366 THz wave is tightly bound on the metal rod surface and located inside the $G_x$ interspace channels, which shows high field confinement in the transverse directions comparing with 0.166 THz. A weak field with loose confinement can be found at 0.522 THz. It exhibits that the higher-order mode (3rd mode) cannot be confined and to be guided by the long MRA. A similar conclusion can be found in reference [19-20], symmetric MRA with 30 layers cannot sustain higher-order TM mode to propagate due to the high impedance. As shown in Figs. 5 (d), the field of 0.170 THz is totally confined inside MRA for a large $G_x$ interspace of 0.815 mm. Noted that larger $G_x$ achieves stronger confinement and higher transmittance of fundamental modes. It agrees well with the trend that has been presented in Fig. 4 that the spectral peak transmittance of the fundamental mode increases for the $G_x$ increment. Compared with $G_x$=0.252 mm, the 0.293 THz field is loosely concentrated inside 0.815 mm-$G_x$ MRA. It means that the larger $G_x$ interspace reduced the 2nd TM mode transmittance. In other words, the largest $G_x$ MRA exhausted the MRA waveguide performance and it is not suitable for 2nd TM mode guiding. To sustain the 2nd TM mode for the longest propagation length, the $G_x$ interspace around 0.252 mm is good for the MRA waveguide. As shown in Figs. 4, both experimental and simulated results demonstrated that the larger $G_x$ interspace leads to an enhanced 3rd TM mode. With the increasing $G_x$, the 3rd TM mode confinement has been improved and thus results in a high transmittance. For example, the transmittance of the 3rd TM mode is larger than 60% when the $G_x$ interspace is 0.815 mm.

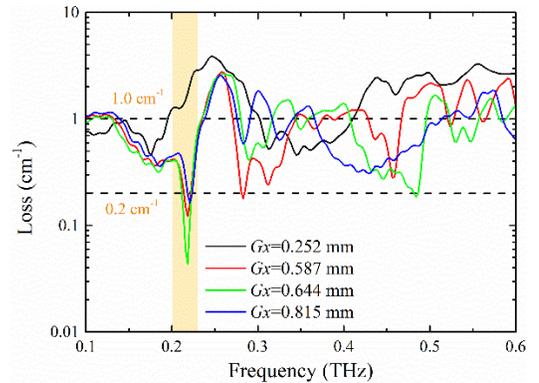

**Fig. 6.** The simulated propagation loss of MPAs with different $G_x$.

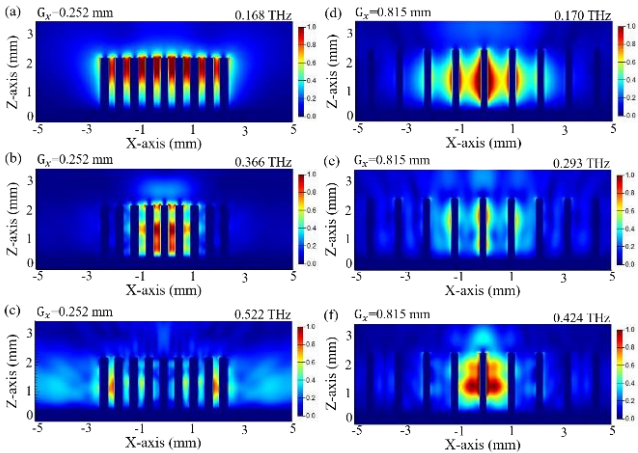

**Fig. 5.** (a), (b) and (c) are the electric field distribution respectively at 0.168, 0.366, and 0.522 THz for 0.252 mm- $G_x$ MRAs. (d), (e) and (f) are

Figure 6 shows the simulated propagation loss of MRAs with different $G_x$. We consider 10 MPA layers (5/15 layers) as one waveguide interval and analyze the waveguide propagation loss of MPAs. The waveguide propagation losses are obtained from the equation $[ln(T_1/T_2)]/[2(L_2−L_1)]$, where $L_{1,2}$ indicates two waveguide lengths for different numbers of MPA layer and $T_{1,2}$ indicates the related transmittance at different numbers of the MPA layer [20]. As shown in Fig. 6, in contrast to the $G_x$=0.252 mm MRA, $G_x$=0.644 mm MRA has the lowest metal rod attenuation and the coefficients are less than 0.2 cm$^{-1}$ at about 0.22 THz. For smaller $G_x$, high waveguide loss of THz waves occurs in the 1st pass-band, corresponding to the fundamental modes. The loss trend of MRA agrees well with that PPWG, which is anti-proportional to the air gap between metal rods [9]. For the $G_x$=0.815 mm MRA, the scattering loss in the frequency of 0.37-0.50 THz is lower than 1 cm$^{-1}$ because of the improved 3rd waveguide TM mode. These results suggest that a 3D printed MRA is a good promising waveguide for THz waves guiding.

In summary, several 3D printed MRAs with various metal rod interspaces ($G_x$) have been investigated in detail by FDTD simulations and THz-TDS measurements. Results show that the spectral positions, bandwidths, and transmittances of TM modes can be manipulated by changing the interspace between metal rods. With the rise of the interspace $G_x$ from 0.252 to 0.815 mm, the fundamental mode shifts to higher frequencies, and its transmittance is increased. For the 2nd TM mode, the transmittance of high-order mode decreases from 0.72 to 0.23. Besides, its frequency is also reduced from 0.366 to 0.293 THz. In contrast to the 2nd mode, the transmittance of the 3rd mode is efficiently improved for the $G_x$ increment. The transmittance of the 3rd TM mode at 0.424 THz is larger than 60% when the $G_x$ interspace is 0.815 mm. Thus, by tuning the air interspace, a mode conversion can be realized in high-frequency bands. The geometry-dependent field performance of the high-order TM mode at the MRA–air interface enables the MRA structure to be flexible for THz applications in communications and sensing.

**Funding.** China Scholarship Council (CSC NO.201606890003). Japan Society for the Promotion of Science (JSPS) (KAKENHI, 16K17525).

**Acknowledgment**. We thank the Ja-Yu Lu (National Cheng Kung University) for providing the measurement system.

**Author contributions.** D. Liu organized and wrote the paper. D. Liu and S. Zhao performed the fabrication and simulations. T. Hattori and B. You provided the fabrication equipment. B. You completed the measurement.

**Disclosures**. The authors declare no conflicts of interest.

### References

1. Y. S. Lee, Principles of terahertz science and technology. Springer Science & Business Media, 2009.
2. S. Atakaramians, S. Afshar, T. M. Monro, et al. "Terahertz dielectric waveguides," Adv. Opt. Photonics, 5(2), 169-215 (2013).
3. K. Nielsen, H. K. Rasmussen, A. J. L. Adam, P. C. M. Planken, O. Bang, and P. U. Jepsen, "Bendable, low-loss Topas fibers for the terahertz frequency range," Opt. Express 17(10), 8592–8601 (2009).
4. S. Atakaramians, S. Afshar Vahid, B. M. Fischer, D. Abbott, and T. M. Monro, "Porous fibers: a novel approach to low loss THz waveguides," Opt. Express 16, 8845–8854 (2008).
5. J. D. Joannopoulos, S. G. Johnson, J. N. Winn, D. M. Robert, Photonic crystal: molding the flow of light. Princeton Univ. Press, Princeton, NJ, 2008.
6. R. Kakimi, M. Fujita, M. Nagai, M. Ashida and T. Nagatsuma, "Capture of a terahertz wave in a photonic-crystal slab," Nat. Photonics 8, 657–663 (2014).
7. D. M. Pozar, Microwave engineering. John Wiley & Sons, 2009.
8. Wang K, Mittleman D M. Metal wires for terahertz wave guiding[J]. Nature, 2004, 432(7015): 376-379.
9. M. Wächter, M. Nagel, and H. Kurz, "Metallic slit waveguide for dispersion-free low-loss terahertz signal transmission," Appl. Phys. Lett. 90(6), 061111 (2007).
10. R. Mendis, D. M. Mittleman, "Comparison of the lowest-order transverse-electric (TE 1) and transverse-magnetic (TEM) modes of the parallel-plate waveguide for terahertz pulse applications," Opt. Express, 17(17), 14839-14850 (2009).
11. D. K. Gramotnev, and S. I. Bozhevolnyi, "Plasmonics beyond the diffraction limit," Nat. Photonics, 4(2), 83–91 (2010).
12. J. A. Schuller, E.S. Barnard, W. Cai, Y.C. Jun, J.S. White, and M. L. Brongersma, "Plasmonics for extreme light concentration and manipulation," Nat. Mater. 9(3), 193-204 (2010).
13. J. B. Pendry, L. Martín-Moreno, and F. J. Garcia-Vidal, "Mimicking surface plasmons with structured surfaces," Science 305, 847 (2004).
14. S. A. Maier and S. R. Andrews, "Terahertz pulse propagation using plasmon-polariton-like surface modes on structured conductive surfaces," Appl. Phys. Lett. 88, 251120 (2006).
15. C. R. Williams, S. R. Andrews, et al. "Highly confined guiding of terahertz surface plasmon polaritons on structured metal surfaces," Nat. Photonic 2, 175 (2008).
16. A. J. Gallant, M. A. Kaliteevski, D. Wood, M., C. Petty, R. A. Abram, S. Brand, G. P. Swift, D. A. Zeze and J. M. Chamberlain, "Passband filters for terahertz radiation based on dual metallic photonic structures," Appl. Phys. Lett. 91, 161115 (2007).
17. A. L. Bingham, D. R. Grischkowsky, "Terahertz 2-D photonic crystal waveguides," IEEE Microw. Wirel. Comp. Lett. 18(7), 428-430 (2008).
18. Kitagawa J, Kodama M, Koya S, et al. THz wave propagation in two-dimensional metallic photonic crystal with mechanically tunable photonic-bands. Optics express, 2012, 20(16): 17271-17280.
19. B. You, D. Liu, T. Hattori, T. A. Liu, and J. Y. Lu, "Investigation of spectral properties and lateral confinement of THz waves on a metal-rod-array-based photonic crystal waveguide," Opt. Express, 26(12), 15570–15584 (2018).
20. D. Liu, J. Y. Lu, B. You, and T. Hattori, "Geometry-dependent modal field properties of metal-rod-array-based terahertz waveguides," OSA Contin. 2(3), 655-666 (2019).
21. H. Wei, K. Li, W. G. Liu, et al. 3D Printing of Free‐Standing Stretchable Electrodes with Tunable Structure and Stretchability. Advanced Engineering Materials, 19(11): 1700341 (2017).
22. W. J. Otter, N. M. Ridler, H. Yasukochi, et al. "3D printed 1.1 THz waveguides." Electron. Lett. 53(7), 471-473 (2017).
23. T. Ma, H. Guerboukha, et al. "3D printed hollow-core terahertz optical waveguides with hyperuniform disordered dielectric reflectors," Adv. Opt. Mater. 6, 1 (2016).
24. J. Li, K. Nallappan, H. Guerboukha, and M. Skorobogatiy, "3D printed hollow core terahertz Bragg waveguides with defect layers for surface sensing applications," Opt. Express 25, 4126–4144 (2017).